\documentclass[aps,prl,twocolumn,nopacs,superscriptaddress,nobibnotes]{revtex4-1}

\usepackage{graphicx}
\usepackage{amsfonts}
\usepackage{amssymb}
\usepackage{multirow}

\usepackage{grffile}
\usepackage{epstopdf}
\usepackage{graphicx}
\usepackage{endnotes}
\usepackage{bm}
\usepackage{epsfig}
\usepackage{amsmath}
\usepackage{color}
\usepackage{textcomp}
\usepackage[colorlinks=true,
            linkcolor=blue,
            urlcolor=blue,
            citecolor=blue,
            hyperfootnotes=true]{hyperref}

\usepackage{array}
\newcolumntype{L}[1]{>{\raggedright\let\newline\\\arraybackslash\hspace{0pt}}m{#1}}
\newcolumntype{C}[1]{>{\centering\let\newline\\\arraybackslash\hspace{0pt}}m{#1}}
\newcolumntype{R}[1]{>{\raggedleft\let\newline\\\arraybackslash\hspace{0pt}}m{#1}}

\newcommand{\kp}{{$k\cdot p$ }}

\newcommand{\K}{\mathcal{K}}
\newcommand{\T}{\mathcal{T}}
\newcommand{\I}{\mathcal{I}}
\newcommand{\Q}{\mathcal{Q}}

\newcommand{\Cz}{\mathcal{C}_{2z}}

\newcommand{\Mz}{\mathcal{M}_z}

\newcommand{\Gz}{\mathcal{G}_z}
\newcommand{\Gy}{\mathcal{G}_y}
\newcommand{\Gx}{\mathcal{G}_x}

\newcommand{\Sy}{\mathcal{S}_y}
\newcommand{\Sx}{\mathcal{S}_x}

\begin{document}

\title{Novel Family of Topological Semimetals with Butterfly-like Nodal Lines}


\author{Xiaoting Zhou}\email[]{physxtzhou@gmail.com}
\affiliation{Department of Physics and Astronomy, California State University, Northridge, CA 91330, USA}
\author{Chuang-Han Hsu}
\affiliation{Department of Electrical and Computer Engineering, Faculty of Engineering, National University of Singapore, Singapore 117583}
\author{Hugo Aramberri}
\affiliation{Department of Physics and Astronomy, California State University, Northridge, CA 91330, USA}
\author{Mikel Iraola}
\affiliation{Donostia International Physics Center, 20018 Donostia-San Sebastian, Spain}
\affiliation{Department of Condensed Matter Physics, University of the Basque Country UPV/EHU, Apartado 644, 48080 Bilbao, Spain}
\author{Cheng-Yi Huang}
\affiliation{Institute of Physics, Academia Sinica, Taipei 11529, Taiwan}
\affiliation{Department of Physics and Astronomy, California State University, Northridge, CA 91330, USA}
\author{Juan L. Ma\~nes}
\affiliation{Department of Condensed Matter Physics, University of the Basque Country UPV/EHU, Apartado 644, 48080 Bilbao, Spain}
\author{Maia G. Vergniory}
\affiliation{Donostia International Physics Center, 20018 Donostia-San Sebastian, Spain}
\affiliation{IKERBASQUE, Basque Foundation for Science, Maria Diaz de Haro 3, 48013 Bilbao, Spain}
\author{Hsin Lin}
\affiliation{Institute of Physics, Academia Sinica, Taipei 11529, Taiwan}
\author{Nicholas Kioussis}\email[]{nick.kioussis@csun.edu}
\affiliation{Department of Physics and Astronomy, California State University, Northridge, CA 91330, USA}


\date{\today}

\begin{abstract}
In recent years, the exotic properties of topological semimetals (TSMs) have attracted great attention and significant efforts have been made in seeking for new topological phases and material realization. In this work, we propose a new family of TSMs which harbors an unprecedented nodal line (NL) landscape consisting of a pair of concentric intersecting coplanar ellipses (CICE) at half-filling. Meanwhile, the CICE at half-filling guarantees the presence of a second pair of CICE beyond half-filling. Both CICEs are linked at four-fold degenerate points (FDPs) at zone boundaries. In addition, we identify the generic criteria for the existence of the CICE in a time reversal invariant {\it spinless} fermion system or a spinfull system with negligible spin-orbital coupling (SOC). Consequently, 9 out of 230 space groups (SGs) are feasible for hosting CICE whose location centers in the first Brillouin zone (BZ) are identified. We provide a simplest model with SG $Pbam$ (No. 55) which exhibits CICE, and the exotic intertwined drumhead surface states, induced by double-band-inversions. Finally, we propose a series of material candidates that host butterfly-like CICE NLs, such as, ZrX$_2$ (X=P,As), GeTe$_5$Tl$_2$, CYB$_2$ and Al$_2$Y$_3$.
\end{abstract}


\maketitle

Topological semimetals (TSMs)~\cite{Fang2016,Weng2016,Armitage2018}
have emerged among the most active frontiers in condensed matter physics in recent years, drawing widespread attention from both the theoretical and the experimental communities. In the noninteracting limit, TSMs describe systems which are characterized by the topologically robust band-crossings manifolds between conduction and valence bands in momentum $k$ space. These mainfolds can be zero-dimensional (0D) nodal points, \textit{e.g.,} three-dimensional (3D) Weyl semimetals (WSMs)~\cite{Wan2011,Burkov2011,Armitage2018} and Dirac semimetals (DSMs)~\cite{Young2012,Wang2012a,Young2015,Chang2017a,Armitage2018,Yu2018}, and one-dimensional (1D) nodal lines/loops, \textit{e.g.,} nodal-line semimetals (NLSMs)~\cite{Burkov2011a,Fang2016}. Around these band-crossings, electron excitations behave drastically differently from the conventional Schr\"{o}dinger fermions in normal metals. For example, the low-energy electrons in 3D DSMs and WSMs resemble the relativistic Dirac and Weyl fermions, making it possible to mimic high-energy physics phenomena. Meanwhile, TSMs are distinguished from normal semimetals by the accompanying topological indices due to the aforementioned manifolds. Moreover, because of these unique electronic features, TSMs present exotic properties in different ways, such as  Fermi arcs~\cite{Xu2015e} and drumhead surface states (SSs)~\cite{Bian2016a} on surfaces of WSMs and NLSMs, respectively, and  novel transport phenomena \textit{e.g.,} the negative magnetoresistance related to the chiral anomaly in both Weyl and Dirac SMs~\cite{Burkov2014a,Zhang2016a,Hu2019}. 

Among TSMs, NLSMs possess the highest variability. NLs can be integrated in various configurations, \textit{e.g.,} a chain link~\cite{Bzdusek2016,Chang2017b,Yan2017,Yu2018}, a Hopf link~\cite{Chang2017b}, and a knot~\cite{Bi2017}, where each of them carries its unique topology. 
Since the essential characteristics, band crossings, of various TSMs are mostly protected by crystalline symmetries, a thorough classification of a particular type of TSMs in all space groups can greatly accelerate the experimental discovery. There exist well-established classifications for DSMs and WSMs~\cite{yang2014,gao2016,wieder2016}, and the triple point semimetals~\cite{zhu2016}. However, for most types of NLSMs proposed today, except the chain link~\cite{Bzdusek2016} and some types of intersecting rings~\cite{gong2018}, the symmetry criteria of the emergence of particular nodal lines remain deficient. 

\begin{figure}
\begin{centering}
\includegraphics[width=\linewidth]{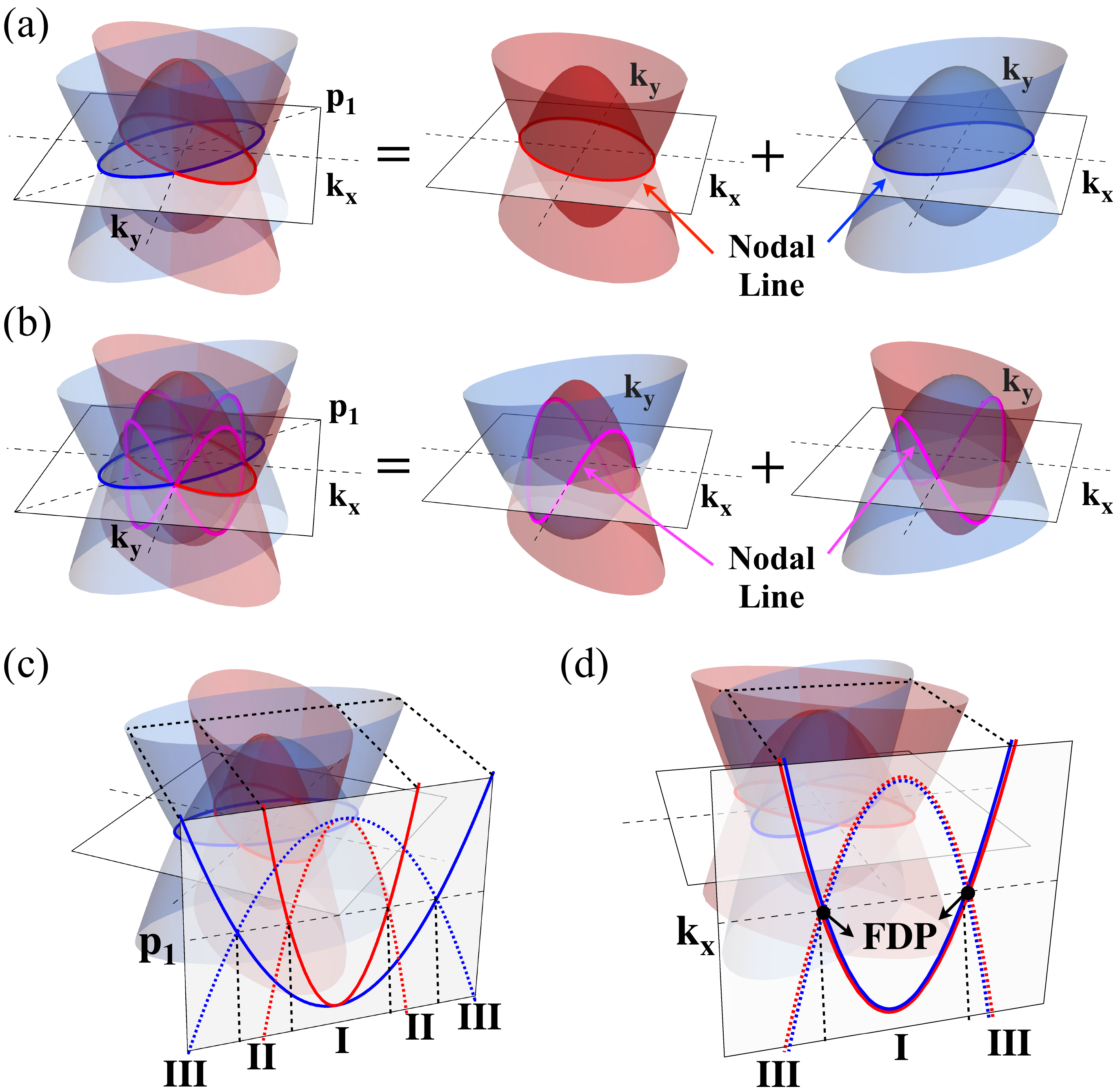}
\par\end{centering}
\centering{}\caption{Schematic band structures showing the mechanism of the formation of the CICE NLs, resulting from the DBI. (a) The half-filling CICE can be decomposed into two ellipses in red and blue, respectively, which are locked by symmetries. Each ellipse is the band-crossing due to a single band inversion and protected by mirror (or glide) symmetry $\Mz$ (or $\Gz$).  (b) The second pair of CICE NLs (labeled in magenta) emerge due to the crossings between the bands denoted in red and blue corresponding to crossings between two occupied (one-quarter filling) or two unoccupied (three-quarter filling) bands. These cross the Fermi surface (FS) leading to the intersection of four NLs at four points on the $k_x$ and $k_y$ axes. Band structures along the paths (c) $p_1$ and (d) $k_x$-axis. The bands labeled by solid and dashed lines carry different eigenvalues of $\Mz$ (or $\Gz$), and opposite parity at the center TRIM point of the CICE. Bands are doubly degenerate along $k_x$-axis (d) and similarly, along $k_y$-axis. I, II and III denote the regions divided by the NLs. }
\label{Fig1}
\end{figure}

In this Letter, we introduce a new type of NLSM in time-reversal invariant {\it spinless} systems, which hosts a butterfly-like nodal-line (NL) consisting of a pair of concentric intersecting coplanar ellipses (CICE) at half-filling residing on a plane in $k$ space, as indicated  by the blue, red concentric ellipses in Fig.\ref{Fig1}(a). Meanwhile, the half-filling CICE consequently guarantees the presence of another pair of CICE formed by band crossings beyond half-filling, which is indicated by the magenta lines in Fig.\ref{Fig1}(b). 
We demonstrate that CICE can be sustained by nonsymmorphic crystalline symmetries including two glide symmetries, and only 9 space groups (SGs) are feasible to host it. These SGs are classified into two categories by their  point group symmetries, which are $D_{2h}$ and $D_{4h}$. Moreover, we provide a tight-binding model for one of the SGs $Pbam$ (No. 55) which exhibits CICE and hosts exotic intertwined drumhead surface states. In the end, five material candidates from these two categories, hosting the proposed CICE, are suggested for further experimental studies.    

{\it Symmetry Criteria and Space Groups--} 
Conceptually, a pair of CICE can be constructed by integrating two NL fermions. As shown in Fig.~\ref{Fig1}(a), CICE denoted by the intersection of the red and blue ellipses on the $k_{z}=0$ plane can be decomposed into two individual NL fermions.
Each NL, the accidental two-fold band-crossings due to the band inversion, is further validated by the inherent crystal symmetry belonging to its parent bands, which is the mirror (or glide) reflection symmetry, $\Mz (\Gz) : (k_x,k_y,k_z)\rightarrow (k_x,k_y,-k_z)$. On the $k_{z}=0$ plane, states carrying different mirror (or glide) eigenvalues forbid their mutual hybridization, thus supporting the NL fermion. Since at each $k$ point $\Mz$ ($\Gz$) only supplies two different eigenvalues, additional symmetry constrains along $k_x$ and $k_y$ axes are demanded to sustain the four-fold degenerate points (FDPs) on the CICE, $i.e.,$ the intersecting points of the two NL fermions, marked in Fig.~\ref{Fig1}(d). 

Therefore, the additional symmetries required along $k_x$ and $k_y$ to guarantee the two-fold degeneracy can be realized by introducing an antiunitary symmetry, $\T \Q$, which combines time-reversal symmetry (TRS) $\T$ and a spatial symmetry $\Q$. Thus, the Kramer-like two-fold degeneracy is enforced at $\T \Q$-invariant points where  $(\T \Q)^2 = -1$. For a {\it spinless} system, $\T^{2}=+1$, which in turn requires that  $\Q$ is a nonsymmorphic symmetry, with eigenvalues of $\pm i$~\cite{Young2015,Wang2016b} at certain points on the boundaries of BZ.  
Consequently, the qualified candidates of $\Q$ for ensuring the degeneracy on the $k_x$ ($k_y$) are, $\Gx$ or $\Sy$ ($\Gy$ or $\Sx$), so the CICE should be centered at $(\pi, \pi, 0\textrm{ or }\pi)$. Here, $\mathcal{S}_{x(y)}=\lbrace \mathcal{C}_{2x(2y)}\vert\bm{t}_{x(y)} \rbrace$ denotes a two-fold screw rotation with respect to the $k_{x(y)}$-axis accompanied by a translation $\bm{t}_{x(y)}=\frac{1}{2}\hat{x}(\hat{y})$, and $\mathcal{G}_{x(y)}=\lbrace \mathcal{M}_{x(y)}\vert\bm{t}_{y(x)} \rbrace$ is a glide symmetry normal to $k_{x(y)}$, which contains a fractional translation $\bm{t}_{y(x)}=\frac{1}{2}\hat{y}(\hat{x})$. To avoid replicas of CICE on other symmetry-related planes, $n$-fold rotation and rotoinversion symmetries with $n>2$ with respect to the $k_{x(y)}$ axes are not allowed.
In addition, each Kramer-paired states should carry the same mirror symmetry eigenvalue of $\Mz$ (or $\Gz$). Furthermore, 
symmetry-enforced degeneracy are not allowed at any generic point of the $k_z =0$ plane other than the $k_{x(y)}$-axis.

\begin{table}
  \begin{center}
    \caption{List of space groups (SGs) and corresponding point groups (PGs) that can host CICE.
    We also list the possible positions of the CICE centers and the axes on which the four-fold 
    degenerate points on the CICE emerge.}
    \label{tab:Table1}
    \setlength{\tabcolsep}{5pt}
    \renewcommand{\arraystretch}{1.2}
    \begin{tabular}{C{0.6cm} L{2.4cm}  L{2.5cm} L{1.7cm}} 
      \hline
      \hline
      \textbf{PGs} & \textbf{SGs} ($\#$) & \textbf{Positions} &\textbf{Axes} \\
      \hline
      \multirow{4}{*}{$D_{2h}$} & $Pbam$ (55) & $(\pi,\pi,0)$, $(\pi,\pi,\pi)$ & \multirow{3}{*} {$\lbrace [100], [010]\rbrace $} \\
      & $Pccn$ (56) & $(\pi,\pi,\pi)$ & \\ 
       & $Pnnm$ (58) & $(\pi,\pi,0)$ &  \\
       & $Pnma$ (62) & $(\pi,0,\pi)$  &  $\lbrace [100], [001]\rbrace $ \\
       \hline
       \multirow{5}{*}{$D_{4h}$} & $P4/mbm$ (127) & $(\pi,\pi,0)$, $(\pi,\pi,\pi)$ & \multirow{5}{*}{$\lbrace [100], [010]\rbrace $} \\
       & $P4/mnc$ (128) & $(\pi,\pi,0)$  &  \\
       & $P4_2/mbc$ (135) & $(\pi,\pi,0)$ & \\
       & $P4_2/mnm$ (136) & $(\pi,\pi,0)$  & \\ 
       & $P4_2/ncm$ (138) & $(\pi,\pi,\pi)$  & \\
      \hline
      \hline
    \end{tabular}
  \end{center}
\end{table}
 
 In summary, the criteria for generating CICE in a {\it spinless} crystal preserving TRS include: (i) the little group of the center of CICE is nonsymmorphic with corresponding point group (PG) $D_{2h}$ or $D_{4h}$; (ii) the crystal contains two glide $\mathcal{G}_{x(y)}$ or screw $\mathcal{S}_{y(x)}$ symmetries with respect to the axes lying on a mirror $\mathcal{M}_z$ or a glide $\mathcal{G}_z$ plane; (iii) at the center of the CICE, $\mathcal{M}_z$ ($\mathcal{G}_z$) should commute with other preserved and required symmetries. According to the above criteria, we have exhaustively scanned all 230 SGs, and determined 9 possible SGs to host CICE. The corresponding positions of the center of the CICE and the corresponding axes of $\mathcal{S}$ or $\mathcal{G}$ are listed in Table.~\ref{tab:Table1}. 

\begin{figure}
\begin{centering}
\includegraphics[width=\linewidth]{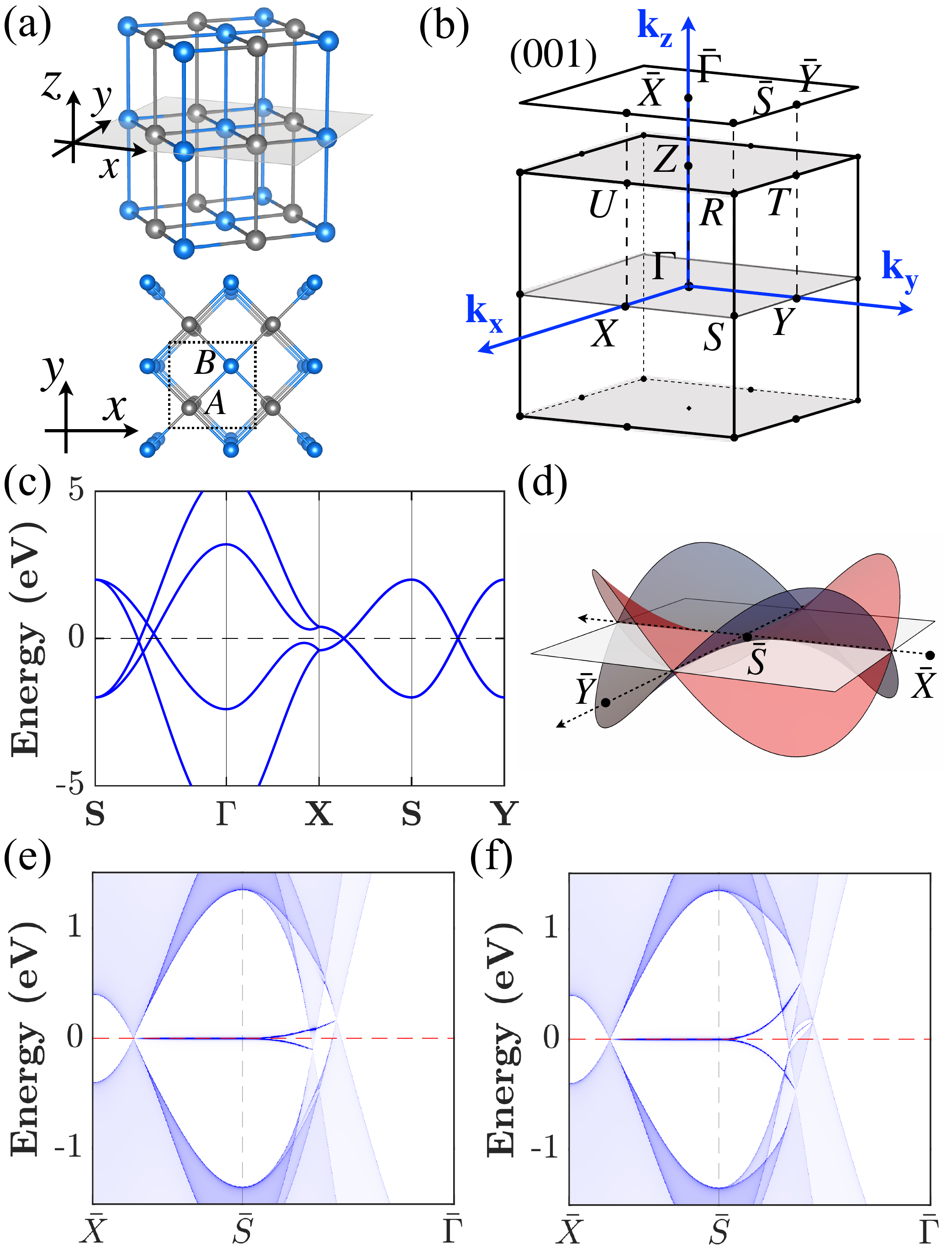}
\par\end{centering}
\centering{}
\caption{(a) Orthorhombic crystal structure of the lattice model in Eq.(\ref{H0}) in SG $Pbam$, consisting of a bipartite lattice with two sublattices A (in blue) and B (in gray). (b)Bulk BZ, the projected $(001)$ surface BZ, and high-symmetry points. (c) Band structures of the model without spin-orbit coupling. (d) Schematic dispersion of the drumhead surface states (DSSs) of the $(001)$ surface stemming from the NLs, with two saddle-like hyperbolic paraboloids intertwined with each other. (e-f) Dispersion along high symmetric directions of two types of DSSs on the (001) surface. (e) Type-I DSSs stem from the CICE FS, while (f) type-II from the NLs labeled in magenta in Fig.\ref{Fig1}(b).}
\label{Fig2}
\end{figure}

{\it Lattice Model and Surface States--} 
To validate the criteria derived above and explore the underlying topology of CICE, we construct a minimal 4-band tight-binding lattice model for the SG $Pbam$ (No. 55). The minimal required symmetries are $\Mz=\lbrace m_{001}\vert 000 \rbrace$, $\Gx=\lbrace m_{100}\vert\frac{1}{2}\frac{1}{2}0 \rbrace$ and $\Gy=\lbrace m_{010}\vert\frac{1}{2}\frac{1}{2}0 \rbrace$ (see Supplementary Material for details). The model is a bipartite lattice, where the sublattices denoted by A (gray) and B (blue) occupy the 2a Wyckoff position at $\mathbf{r}_A=(0, 0, 0)$ and $\mathbf{r}_B=(\frac{1}{2}, \frac{1}{2}, 0)$ in a unit cell (see Fig.~\ref{Fig2} (a) for the structure). Each sublattice contains two orbitals, $p_z$ and $d_{xy}$ described by the Pauli matrix $\bm{\sigma}$, and $\bm{\tau}$ for the A and B sublattices. 
For a {\it spinless} system, employing the basis $\Psi = (p_{z}^A, d_{xy}^A, p_{z}^B, d_{xy}^B)^T$ the symmetry-constrained tight-binding Hamiltonian is of the form, 
\begin{eqnarray}
H_0(\mathbf{k}) &=& [(\alpha \textrm{cos}k_x +\beta \textrm{cos}k_y + \gamma \textrm{cos}k_z) +\delta_0]\tau_{0} \sigma_{3} \nonumber \\
& + & \textrm{cos}\frac{k_x}{2} \textrm{cos}\frac{k_y}{2} \textrm{cos}{k_z} (\lambda_{10}\tau_{1} \sigma_{0} + \lambda_{13} \tau_{1} \sigma_{3})\nonumber \\
& + & \textrm{sin}{k_z} (\lambda_{32} \tau_{3} \sigma_{2}) \nonumber \\
& + & \textrm{sin}\frac{k_x}{2} \textrm{sin}\frac{k_y}{2} \textrm{sin}{k_z}(\lambda_{12}  \tau_{1} \sigma_{2}), 
\label{H0}
\end{eqnarray}
where $\alpha,\beta,\gamma,\delta_0,$ and $\lambda_{ij}$ are constants.

Since the CICE can emerge on the mirror plane (gray shaded area in Fig.~\ref{Fig2}(b)) centered at the high symmetry $k$ point $S=(\pi,\pi,0)$ [$R=(\pi,\pi,\pi)$] (Fig.~\ref{Fig2}(b)), we derive the effective \kp Hamiltonian around the S(R) point,
\begin{eqnarray}
H_{S(R)}(\mathbf{q}) & = & \frac{1}{2}(\alpha q^2_{x} +\beta q^2_{y} \mp\gamma q^2_{z} + 2 \delta_{S(R)}) \tau_{0} \sigma_{3} \nonumber \\
& + & \frac{1}{4} q_x q_y (\lambda_{10} \tau_{1} \sigma_{0} + \lambda_{13} \tau_{1} \sigma_{3}) \nonumber \\
& + & q_z (\lambda_{12} \tau_{1} \sigma_{2} + \lambda_{32} \tau_{1} \sigma_{2}).
\label{kpmodel} 
\end{eqnarray}
At $q_z =0$, the Hamiltonian is diagonalized as $E_{S(R)} (q_x, q_y, 0) = \textrm{diag}(E_+^{A+B}, E_-^{A+B}, E_+^{A-B}, E_-^{A-B})$ on the basis $\Psi' =(p_z^{A+B}, d_{xy}^{A+B}, p_z^{A-B}, d_{xy}^{A-B})^T$, where $|\varphi^{A\pm B}\rangle = \frac{1}{\sqrt{2}} (|\varphi^A\rangle \pm |\varphi^B\rangle) $ ($\varphi = p_z, d_{xy}$) denote the bonding/antibonding states of the relevant orbitals. 

If $|\lambda_{13}| > |\lambda_{10}|$, each ellipse in the half-filling CICE is the line crossing between the two bands $E_+^{A+B}$ ($E_+^{A-B}$) and $E_-^{A+B}$ ($E_-^{A-B}$) respectively, while the ellipses in the second pair are given by the crossings between  $E_+^{A+B}$ ($E_+^{A-B}$) and $E_-^{A-B}$ ($E_-^{A+B}$).  We refer to this type of CICE as type-I NL. Otherwise, if $|\lambda_{13}| < |\lambda_{10}|$, the first and second pairs of CICE are exchanged and we have type-II NL. Thus, the corresponding NLs for the half-filling CICE can be obtained by solving the equations,
\begin{eqnarray}\label{eq:ellipse}
|\lambda_{13}| > |\lambda_{10}|:\;\; \alpha q^2_{x} + \beta q^2_{y} \pm \frac{1}{2} \lambda_{13} q_x q_y + 2\delta_{S(R)} &=& 0, \ \ \  \label{b-b} \\
|\lambda_{13}| < |\lambda_{10}|:\;\; \alpha q^2_{x} + \beta q^2_{y} \pm \frac{1}{2} \lambda_{10} q_x q_y + 2\delta_{S(R)} &=& 0.  \ \ \  \label{b-antib}
\end{eqnarray}
After further analyses, we find that when the condition
\begin{eqnarray}\label{condition}
\lbrace \alpha \delta_{S (R)} <0 \ \cap \ \alpha \beta > 0 \ \cap \ \alpha \neq \beta \rbrace  
\end{eqnarray}
is satisfied, where $\delta_{S,R} = \delta_0 - (\alpha +\beta \mp \gamma)$, the terms in the first line of Eq.~(\ref{H0}) describe two concentric elliptic NLs with double band inversions at the S (R) point (Fig.~\ref{Fig1}(d)). The terms in the second line in Eq.~(\ref{H0}) adjust the anisotropy of each NL, resulting in two twisted elliptic NLs (see dispersion along $k_{x}=k_{y}$ in Fig~\ref{Fig2}(c), where the band width differs in the two original elliptic NLs). The angles of the elliptic NLs with respect to the $k_x$-axis are determined via $\theta_{\pm} = \pm \frac{1}{2}\textrm{arctan} \frac{\lambda}{2(\alpha -\beta)}$, where $\lambda = \textrm{max} \lbrace |\lambda_{10}|, |\lambda_{13}| \rbrace$.

To explore the unique topological properties of CICE NL, the model parameters are tuned to allow the system to host CICE centered at the $S$ point and to have no additional band inversions at other time-reversal-invariant momentum points (TRIM). The corresponding band structure is shown in Fig.~\ref{Fig2}(c), in which the distinctive features of CICE can be recognized by comparing the bands along $S-\Gamma$ and $S-X$ (or $S-Y$).  Since CICE are composed by two NLs, and the essential $\Gx$ and $\Gy$ symmetries are preserved on (001) surface, we anticipate to observe two intertwined drumhead surface states (DSSs)~\cite{Burkov2011a}. The DSS, shown schematically in Fig.~\ref{Fig2}(d), can be described by the \kp Hamiltonian around the $\bar{S}$ point,
\begin{eqnarray}
H_{DSS} (q_x, q_y)  &=& q_x q_y (a_3 \mu_3 + a_1 \mu_1) ,
\label{DSSs}
\end{eqnarray}
where $\mu_{1,2,3}$ are Pauli matrices acting in orbital space, and $a_{1,3}$ are real constants. Two saddle-like hyperbolic paraboloids (red and gray surfaces) are intertwined with each other, resulting in the flat and doubly degenerate bands along $\bar{S}-\bar{X}$ and $\bar{S}-\bar{Y}$ respectively, which are enforced by $\Gx$ and $\Gy$ symmetries combined with $\T$. We would like to emphasize that these remarkable features exhibited by the new DSSs allow them to provide a great platform for studies of exotic emergent phenomena. 

The calculated $(001)$ surface band structure along $\bar{X}-\bar{S}-\bar{\Gamma}$ for $\vert\lambda_{10}\vert  < \vert\lambda_{13}\vert$  is shown in Fig.~\ref{Fig2}(e). Intriguingly, we notice that another type of DSSs, shown in Fig.~\ref{Fig2}(f), can be realized when $\vert\lambda_{10}\vert >\vert\lambda_{13}\vert$ with all remaining parameters unchanged. We refer to the two different types of DSSs as type-I/type-II for the former/latter case. Type-II DSSs can be comprehended from the way one proceeds to decompose the second pair of CICE into two single NLs. As shown in Fig.~\ref{Fig1}(b) the NLs (shown by magenta color) are allowed by the same band configurations with swapped conduction bands in comparison to the configurations of Fig.~\ref{Fig1} (a). In contrast to the NLs of the half-filling CICE, the NLs in magenta are due to the crossings between two occupied bands (one-quarter filling) and two unoccupied bands (three-quarter filling), respectively, and hence might be irrelevant for electron excitation at half-filling system. However, the CICE TSM introduces another possibility. 
As one cannot distinguish whether the FDPs of CICE belong to the half-filling CICE or the second pair of CICE, both of them can provide topological DSSs on an equal footing due to the inherent band inversion. Even though both types of NL contribute to the DSS on the (001) surface, $\Gx$ and $\Gy$ permit solely one pair of DSSs, forcing in turn the other pair merge into the bulk states. Consequently, the DSS of CICE-NL in the {\it spinless} case may appear in either way depending on the coupling parameter details. 

\begin{figure}
\begin{centering}
\includegraphics[width=\linewidth]{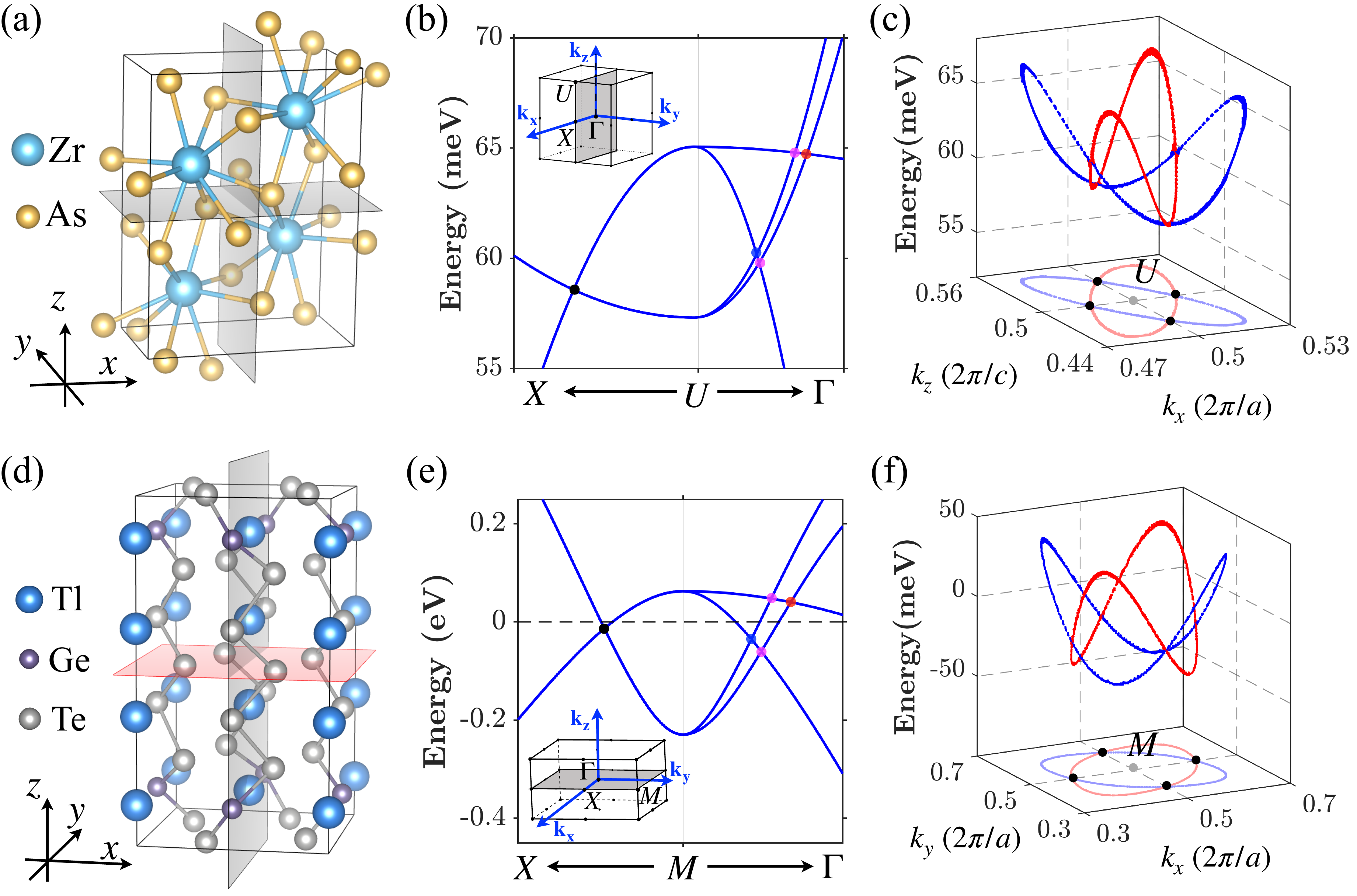}
\par\end{centering}
\centering{}
	\caption{Crystal structure of bulk (a) ZrAs$_2$ and (d)  GeTe$_{5}$Tl$_{2}$. Band structure of (b)  ZrAs$_2$ and (e) GeTe$_{5}$Tl$_{2}$ close to the Fermi energy without SOC along the symmetry lines in the BZ shown in the insets where high-symmetry points are marked. The relevant crossings on $U$ ($M$)-$\Gamma$ for the half-filling and the second pair of CICE are indicated by $[$red, blue$]$ and magenta dots, respectively. Energy-momentum spread for the half-filing CICE nodal lines in the BZ for (c) ZrAs$_2$ centered at U point on the (010) plane (grey shaded) and (f) GeTe$_{5}$Tl$_{2}$ at M point on the (001) plane (grey shaded). The black dots shown on the projected CICEs in the $k$ space are the FDPs and one of them is indicated in the band along $X$-$U$ ($M$) of (b) and (e). As suggested by the crossings in (b) and (e), the energy spreads for the second pair of CICE are very similar.}
	\label{Fig3}
\end{figure} 

{\it Material Candidates--} 
We propose a series of compounds as material candidates for the experimental realization of this new type of TSMs that host butterfly-like CICE NLs, such as ZrX$_2$ (X = As, P) with CICE centered at TRIM point  U ($\pi,\pi,0$), as well as GeTe$_{5}$Tl$_{2}$, CYB$_2$ and Al$_2$Y$_3$ at M ($\pi,\pi,0$), respectively. Here we take the ZrAs$_2$ and GeTe$_{5}$Tl$_{2}$ as representatives. The equilibrium lattice constants and electronic structure of both compounds were determined by first principles density functional theory (DFT) calculations using the VASP~\cite{VASP} and WIEN2k~\cite{blaha2020}~packages (CYB$_2$ and Al$_2$Y$_3$ were found using the \emph{Advanced Search Tools} of \url{https://www.topologicalquantumchemistry.org/}\cite{TQC,MatCat}, see Supplementary Materials for details).

The crystal structure of ZrAs$_2$, is orthorhombic with SG $Pnma$ (No. 62) and is displayed in Fig.~\ref{Fig3} (a). The calculated lattice parameters $a=6.847$~\AA, $b=3.718$~\AA\  and $c=9.123$~\AA\ are in agreement with the experimental ones \cite{ZrAs2exp,ZrAs2exp2}. The band structure without SOC close to the $U=(\pi,0,\pi)$ point of the BZ is shown in Fig.~\ref{Fig3} (b) along two high-symmetry lines ($X-U$ and $U-\Gamma$).  The second material, GeTe$_{5}$Tl$_{2}$, has tetragonal structure with SG $P4/mbm$ (No. 127) and the crystal structure is depicted in Fig.~\ref{Fig3} (d). In the DFT calculations the experimental structure of GeTe$_{5}$Tl$_{2}$~\cite{Marsh1990,AbbaToure1990} is applied, and the band structures along two essential high symmetry $k$ paths $M\rightarrow X$ and $M\rightarrow \Gamma$ are shown in Fig.~\ref{Fig3} (d), which reveal  band crossings alike those observed in ZrAs$_{2}$. 

In order to corroborate the CICE NLs in ZrAs$_2$ around $U$ we used the Bloch functions obtained with DFT to construct a Wannier-function based model employing the Wannier90 package\cite{Wannier90}.
The model reproduces the bands around the Fermi level, allowing the scan of band crossings in the BZ  more efficiently than direct DFT calculations. As shown in Fig.~\ref{Fig3} (c) the nodal points around $U$ point form a butterfly-like CICE with a small energy dispersion. In the case of GeTe$_{5}$Tl$_{2}$, the band crossings, shown in Fig~\ref{Fig3} (f), occur closer to the $M$ point, yielding a smaller area enclosed by the CICE and a lower energy dispersion.

{\it Conclusion--} 
In summary, we have proposed a new type of TSMs which unveil intriguing butterfly-like CICE NDLs. We have derived the symmetry criteria to generate the CICE, identified the 9 SGs which can host such complex NDLs and determined the positions of the CICE centers in the BZ for each SG. For one of the SGs $Pbam$ (No. 55) and for {\it spinless} fermions we have introduced a model which hosts CICE and supports the intriguing intertwined drumhead surface states. Finally, we have predicted candidate materials which can host such exotic NL landscapes.

{\it Acknowledgments--}
The work at CSUN was supported by NSF-Partnership in Research and Education in Materials (PREM) Grant No. DMR-1828019. The work of J.L.M. has been supported by Spanish Science Ministry grant PGC2018-094626-B-C21 (MCIU/AEI/FEDER, EU) and Basque Government grant IT979-16. M.G.V. thanks support from DFG INCIEN2019-000356 from Gipuzkoako Foru Aldundia. H.L. acknowledges the support by the Ministry of Science and Technology (MOST) in Taiwan under grant number MOST 109-2112-M-001-014-MY3.

\bibliographystyle{apsrev4-1}
%
%

\clearpage
\widetext
\setcounter{equation}{0}
\setcounter{figure}{0}
\setcounter{table}{0}
\renewcommand{\theequation}{S\arabic{equation}}
\renewcommand{\thefigure}{S\arabic{figure}}
\renewcommand{\thetable}{S\arabic{table}}
\renewcommand{\bibnumfmt}[1]{[S#1]}
\renewcommand{\citenumfont}[1]{S#1}
\newcommand{\bk}{\boldsymbol\kappa}

\newcommand{\SI}{Supplementary Material}
\newcommand{\beginsupplement}{%
  \setcounter{equation}{0}
  \renewcommand{\theequation}{S\arabic{equation}}%
  \setcounter{table}{0}
  \renewcommand{\thetable}{S\arabic{table}}%
  \setcounter{figure}{0}
  \renewcommand{\thefigure}{S\arabic{figure}}%
  \setcounter{section}{0}
  \renewcommand{\thesection}{S\Roman{section}}%
  \setcounter{subsection}{0}
  \renewcommand{\thesubsection}{S\Roman{section}.\Alph{subsection}}%
}

\section{Supplemental material}

\subsection{I. Symmetry Constraints and Tight-Binding Model}
As mentioned in the main text, the CICE can be constructed from a bipartite lattice, with two sublattices denoted by A and B occupying the 2a Wyckoff position at $\mathbf{r}_A=(0, 0, 0)$ and $\mathbf{r}_B=(\frac{1}{2}, \frac{1}{2}, 0)$ in a unit cell. 
For spinless systems with SG $Pbam$ (No. 55), the symmetry constraints at TRIM points take the form
\begin{eqnarray}
\Gamma=(0,0,0) \cup Z=(0,0,\pi) &:& \ \ 
\T = \K, \ \ \I = -\sigma_{3}, \ \ \ \Mz = -\sigma_{3},  \ \ \Gx = e^{-iq_y/2} \tau_{1} \sigma_{3}, \ \ \Gy = e^{-iq_x/2} \tau_{1}\sigma_{3}. \label{sysc} \\
S=(\pi,\pi, 0) \cup R=(\pi, \pi, \pi) &:& \ \ 
\T = \K, \ \ \I = -\sigma_{3}, \ \ \ \Mz = -\sigma_{3}, \ \ \Gx = -e^{-iq_y/2} i\tau_{2} \sigma_{3}, \ \ \Gy = -e^{-iq_x/2} i\tau_{2}\sigma_{3}. \nonumber \\
X=(\pi,0,0) \cup U=(\pi,0,\pi) &:& \ \ 
\T = \tau_{3} \K, \ \ \I = -\tau_{3}\sigma_{3}, \ \ \ \Mz = -\sigma_{3}, \ \ \Gx = e^{-iq_y/2} \tau_{2} \sigma_{3}, \ \ \Gy = -e^{-iq_x/2} i\tau_{1}\sigma_{3}. \nonumber \\
Y=(0, \pi, 0) \cup T=(0, \pi, \pi) &:& \ \ 
\T = \tau_{3} \K, \ \ \I = -\tau_{3}\sigma_{3}, \ \ \ \Mz = -\sigma_{3}, \ \ \Gx = -e^{-iq_y/2} i\tau_{1} \sigma_{3}, \ \ \Gy = e^{-iq_x/2} \tau_{2}\sigma_{3}. \nonumber
\end{eqnarray}

where $q_{x,y}$ is the $k$ vector from the corresponding TRIM point, Pauli matrices $\tau$ and $\sigma$ are used for the sublattice and orbital space respectively, and $\K$ is the complex conjugate operator. Given these constrains, the minimal tight-binding model reads 
\begin{eqnarray}
H_0(\mathbf{k}) &=& [(\alpha \textrm{cos}k_x +\beta \textrm{cos}k_y + \gamma \textrm{cos}k_z) +\delta_0]\tau_{0} \sigma_{3} \nonumber \\
& + & \textrm{cos}\frac{k_x}{2} \textrm{cos}\frac{k_y}{2} \textrm{cos}{k_z} (\lambda_{10}\tau_{1} \sigma_{0} + \lambda_{13} \tau_{1} \sigma_{3})\nonumber \\
& + & \textrm{sin}{k_z} (\lambda_{32} \tau_{3} \sigma_{2}) \nonumber \\
& + & \textrm{sin}\frac{k_x}{2} \textrm{sin}\frac{k_y}{2} \textrm{sin}{k_z}(\lambda_{12}  \tau_{1} \sigma_{2}), 
\label{H0si}
\end{eqnarray}
which is Eq. (1)  in the main text. The basis considered here is $\Psi = (p_{z}^A, d_{xy}^A, p_{z}^B, d_{xy}^B)^T$. According to the symmetry constraints given in Eq.~\ref{sysc}, at the S (R) point the eigenvalues of the two glide-mirror symmetries are $\pm i$, and $\Mz$ commutes with all the other spatial symmetries. Thus, the emergence criteria of the CICE discussed in the main text are satisfied at the S (R) point. 

To explore the unique topological properties of CICE NL, the system is designed to host CICE centered at the $S$ point and to have no additional band inversions at other time-reversal-invariant momentum points (TRIM). Thus, in addition to $\lbrace \alpha \delta_{S} <0 \ \cap \ \alpha \beta > 0 \ \cap \ \alpha \neq \beta \rbrace $, the model parameters should be tuned to satisfy the following conditions:
\begin{eqnarray}
&\lbrace  \delta_S  < 0  \cap \delta_R  > 0  \cap  \delta_{\Gamma} > max (|\lambda_{10}| , |\lambda_{13}| )  \cap \   \delta_{Z} > max (|\lambda_{10}| , |\lambda_{13}| ) \rbrace;\\ \nonumber
&\textrm{or}& \\ \nonumber
&\lbrace  \delta_S  > 0 \cap \delta_R  < 0 \cap  \delta_{\Gamma} < min (-|\lambda_{10}| , -|\lambda_{13}| )  \cap \   \delta_{Z} < min (-|\lambda_{10}| , -|\lambda_{13}| ) \rbrace;\\ \nonumber
\end{eqnarray}
where
\begin{eqnarray}
\delta_{\Gamma} = \delta_0 + (\alpha +\beta +\gamma), \ \ \
\delta_Z = \delta_0 + (\alpha +\beta -\gamma), \\ \nonumber
\delta_S = \delta_0 - (\alpha +\beta -\gamma), \ \ \ 
\delta_R = \delta_0 - (\alpha +\beta +\gamma), \\ \nonumber
\end{eqnarray}


\subsection{II. Surface States}
In this section, we construct the \kp Hamiltonians for (001) surface states, to describe the drumhead surface states (DSSs) stemming from the CICE NLs in the absence of spin-orbit coupling (SOC), and the topological surface states (TSSs) when SOC is considered.

For a system in space group $Pbam$ (No.55), the (001) surface is  stabilized by the two-dimensional (2D) wallpaper group $P2gg$, which contains one rotational symmetry along the $z$-axis, $\Cz=\lbrace 2_{001}|000 \rbrace$, and two glide-mirror symmetries $\Gx=\lbrace m_{100}|\frac{1}{2} \frac{1}{2} 0\rbrace$ and  $\Gy=\lbrace m_{010}|\frac{1}{2} \frac{1}{2} 0\rbrace$. Beyond these crystalline symmetries, time-reversal symmetry (TRS) $\T$ is also preserved. Since the CICE can only  emerge at S or R point (see Fig.2(b) in the main text), which projects to $\bar{S}$ on the (001)-surface, the corresponding \kp surface Hamiltonian should be constructed around $\bar{S}$.

In the absence of SOC, the DSSs from the CICE are composed by two copies of DSS originating from each ellipse NL, and therefore a two-band model is required. At the $\bar{S}$ point, where  $\T = \K$ since $\T$ should commute with all spatial symmetries, and both $\Gx$ and $\Gy$ should take the eigenvalues $\pm i$, the symmetry constraints can be determined as $\Gx = i\mu_2$, $\Gy = i\mu_2$, and accordingly, $\Cz = \Gx*\Gy =-\mu_0$. Here, $\mu_{1,2,3}$ and $\mu_0$ are respectively the Pauli matrices and  the identity matrix acting in orbital space. As a result, the Hamiltonian for the DSSs from the CICE on the  (001) surface is 
\begin{eqnarray}
H_{DSS} (q_x, q_y)  &=& q_x q_y (a_3 \mu_3 + a_1 \mu_1)
\label{DSSs}
\end{eqnarray}
where $a_{1,3}$ are real constants.

\subsection{III. Details of the DFT calculations}
In the density functional theory (DFT) calculations of ZrAs$_2$ and ZrP$_2$ the Perdew-Burke-Ernzherhof (PBE)~\cite{PBE} implementation of the generalized gradient approximation (GGA) was used for the exchange-correlation functional.
A plane-wave basis with an energy cutoff of 340 eV was employed in all calculations. The BZ was sampled with an 8$\times$15$\times$6 Monkhorst-Pack grid~\cite{Monkhorstpack}. The systems were allowed to fully relax until residual atomic forces became smaller than 0.01 eV/\AA.

\subsection{IV. Details of the Wannier function-based model}
A Wannier function-based model for ZrAs$_2$ and ZrP$_2$ was obtained using the DFT results as starting point.
The Wannier90 package along with its interface with the VASP package were employed for this purpose~\cite{Wannier90si}
The upper limit of the frozen energy window was set to 2.5 eV above the Fermi level of each compound. The model perfectly reproduces the DFT bands up to that energy, which is well above the studied band crossings. 
We chose the following atomic orbitals as the starting guess for the projection of the Bloch states onto
 Wannier functions: $s$, $p$ and $d$ orbitals for Zr and $s$ and $p$ orbitals for As or P.
\begin{figure}[h!]
\centering
\includegraphics[width=0.8\textwidth]{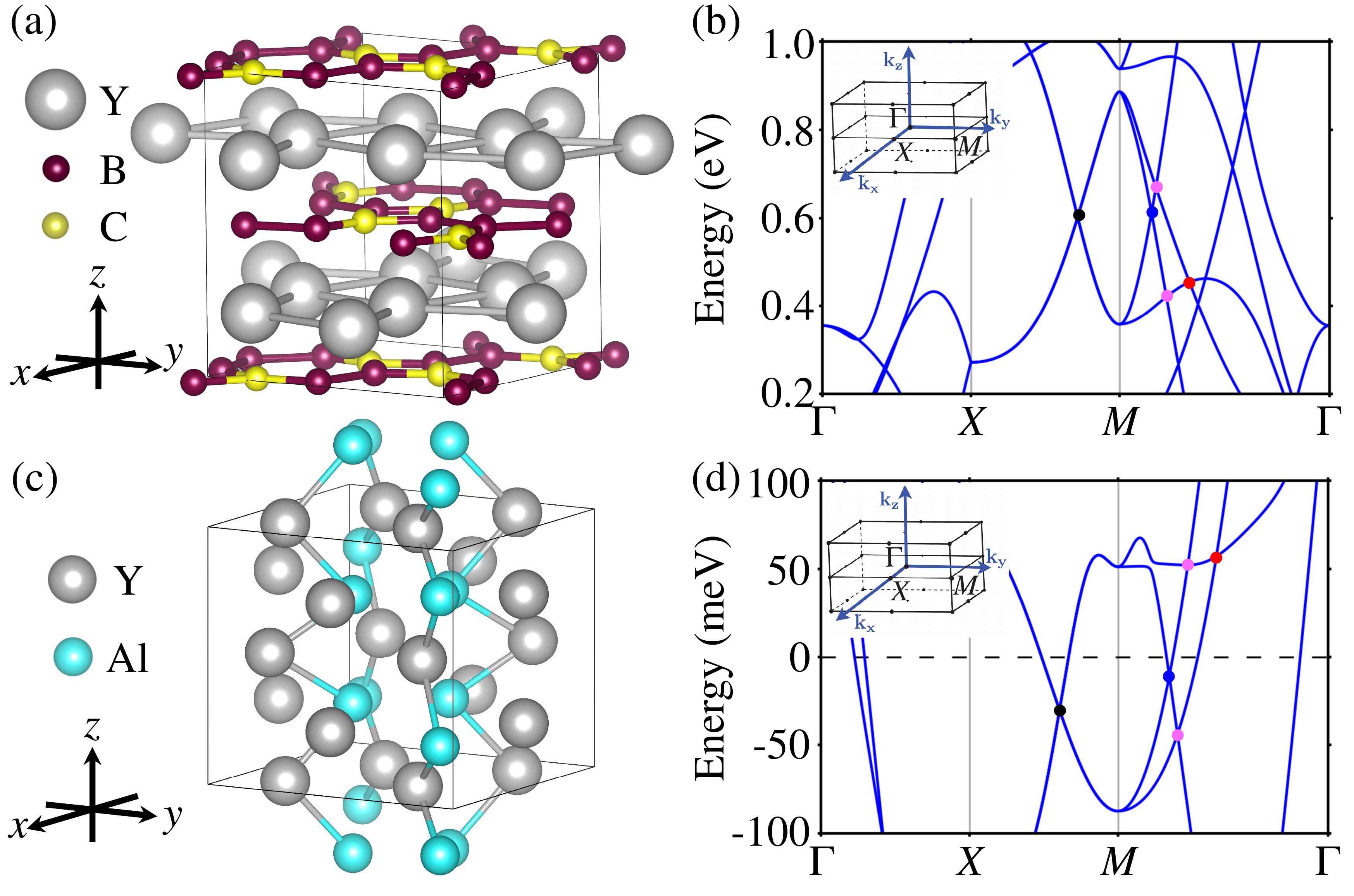}
\caption{Crystal structure of bulk (a) CYB$_2$ with tetragonal space group (SG) $P4_2/mbc$ (No.135), (c) Al$_2$Y$_3$ with tetragonal SG $P4/mbm$ (No. 136). Band structure of (b)  CYB$_2$ and (d) Al$_2$Y$_3$ close to the Fermi energy without spin-orbital coupling (SOC) along the symmetry lines in the Brillouin zone shown in the insets where high-symmetry points are marked. The relevant crossings on $M$-$\Gamma$ for the half-filling and the second pair of CICE are indicated by $[$red, blue$]$ and magenta dots, respectively. The black dots along $X$-$M$ are the four-fold degenerate points (FDPs). }
\label{fig:good-bands}
\end{figure}
\subsection{V. Material candidates in $P4_2/mbc$ (No.135) and $P4_2/mnm$ (No. 136)} We propose CYB$_2$\cite{CYB2exp} and Al$_2$Y$_3$\cite{Al2Y3exp} as materials candidates to host CICE. The candidates were identified using the \emph{Advanced Search Tools} of \url{https://www.topologicalquantumchemistry.org/}\cite{TQCsi,MatCatsi}.
The crystal structure of CYB$_2$ is tetragonal with SG P4$_2$/mbc (No.135) and shows a layered structured of Y layers intercalated with B and C nets. Al$_2$Y$_3$ crystal structured belongs to P4$_2$/mnm (No. 136) tetragonal SG. Figs. \ref{fig:good-bands} (a) - (c)  and (b) - (d)  show the band structure calculation and crystal structure of CYB$_2$ and Al$_2$Y$_3$ respectively. The band structures include two essential high symmetry paths $X$ $\rightarrow$ $\Gamma$ $\rightarrow$ $M$, where the crossings of CICE are highlighted by black dots. 
These bands structures were calculated using VASP with the the modified Becke-Johnson exchange potential in combination with GGA \cite{MBJ1,MBJ2}. The BZ was sampled with an 7x7x9 Monkhorst-Pack grid and an energy cut off of 520 eV was used.




\bibliographystyle{apsrev4-1}
%

%

\end{document}